\documentclass[12pt]{article}
 \usepackage{color}
\usepackage[usenames,dvipsnames]{xcolor}
\usepackage[utf8]{inputenc}
\usepackage[english]{babel}
\usepackage{color}
\usepackage{amsmath}
\usepackage{amsfonts}
\usepackage{amssymb}
\usepackage[margin=0.5cm]{caption}
\usepackage{graphicx}
\usepackage{verbatim}
\usepackage{braket} 
\usepackage{subfig}
\usepackage{float}
\usepackage{simplewick}

 \usepackage[nottoc]{tocbibind}
\usepackage[colorlinks=true,linkcolor=blue,urlcolor=magenta, citecolor=blue]{hyperref}
\usepackage{fullpage}
\usepackage{amssymb,graphicx}
\usepackage{amsmath}
\usepackage[margin=0.5cm]{caption}
\usepackage{hyperref} 
\usepackage{cite}
\usepackage{upgreek}

\def\d{\partial}
\def\l{\left(}
\def\r{\right)}

\newcommand{\be}{\begin{equation}}
\newcommand{\ee}{\end{equation}}
\newcommand{\bea}{\begin{eqnarray}}
\newcommand{\eea}{\end{eqnarray}}
\newcommand{\bg}{\begin{gather}}
\newcommand{\eg}{\end{gather}}
\newcommand{\bseq}{\begin{subequations}}
\newcommand{\eseq}{\end{subequations}}

\renewcommand{\ln}{\mathop{\rm ln}\nolimits}

\begin{document}

\baselineskip=15.5pt
\begin{titlepage}
	\begin{center}
		{\Large\bf  Undressing Confining Flux Tubes with  $T\bar T$}\\
		\vspace{0.5cm}
		{ \large
			Chang Chen$^{a}$, Peter Conkey$^{a}$, Sergei Dubovsky$^{a}$,\\
			\vspace{0.2cm} and Guzm\'an Hern\'andez-Chifflet$^{a,b}$ 
		}\\
		\vspace{.45cm}
		{\small  \textit{  $^a$Center for Cosmology and Particle Physics,\\ Department of Physics,
				New York University\\
				New York, NY, 10003, USA}}\\ 
		\vspace{.1cm}
		\vspace{.25cm}         
		{\small  \textit{   $^b$Instituto de F\'isica, Facultad de Ingenier\'ia,\\ Universidad de la Rep\'ublica,\\
				Montevideo, 11300, Uruguay}}\\ 
	\end{center}
	\begin{center}
		\begin{abstract}
		 Lattice QCD simulations provide crucial information about the worldsheet dynamics of confining strings (flux tubes). An accurate extraction of the worldsheet $S$-matrix from lattice spectra requires accounting for polarization effects.  Approximate integrability of the low energy worldsheet theory makes it possible to apply the Thermodynamic Bethe Ansatz to incorporate polarization effects at all orders in the number of windings and at the leading order in the derivative expansion. However, a
		 systematic application of this technique in the presence of non-integrable effects and for multiparticle states becomes increasingly  challenging. We point out that a recently understood equivalence between gravitational dressing and $T\bar{T}$ deformation provides a fully systematic and straightforward recipe to incorporate the leading polarization effects in the presence of an arbitrary inelasticity and for any number of particles. We illustrate this technique with several examples.
		 
		\end{abstract}
	\end{center}
\end{titlepage}
\newpage

\section{Introduction and Summary}
Yang--Mills theory is expected to admit a weakly coupled string description in the limit when the number of colors $N_c$ is large \cite{tHooft:1973alw}. Constructing the corresponding free string theory
has proven to be very challenging, even though a remarkable progress has already been achieved for certain superconformal theories as a result of a happy marriage of the AdS/CFT \cite{Maldacena:1997re,Gubser:1998bc,Witten:1998qj}
and integrability \cite{Beisert:2010jr}.

The first natural step towards making progress for  confining theories, such as pure glue, is to understand the dynamics on the worldsheet
of a single long confining string (a flux tube).
Lattice simulations  \cite{Athenodorou:2010cs,Athenodorou:2011rx,Athenodorou:2016kpd,Athenodorou:2016ebg,Athenodorou:2017cmw} (see \cite{Teper:2009uf,Lucini:2012gg} for reviews)
allow to extract the finite volume spectrum of the worldsheet theory by measuring two point correlation functions of (deformed) Polyakov loops. 

Much of the recent progress in the study of confining strings  is related to a simple realization \cite{Dubovsky:2013gi,Dubovsky:2014fma} that this setup provides a version of the classic lattice QCD problem \cite{Luscher:1990ux} ---
extraction of scattering amplitudes from the finite volume spectrum. The present version of the problem exhibits a couple of peculiar features which make some of its aspects much easier and others quite a bit harder compared to  more conventional settings, such as the extraction of pion scattering amplitudes from lattice data (see \cite{Briceno:2017max} for a recent review). 

An obvious simplification is that for confining strings one is always interested in the two-dimensional scattering amplitudes on the worldsheet, independently of the number of dimensions $D$ where an underlying gauge theory lives in.  
Hence, this setting provides an ideal testing ground for proposals (such as  \cite{Hansen:2014eka}) to extend the L\"uscher quantization condition to multiparticle scattering. Note that in the case at hand the two-dimensional problem is not just a toy model, but rather has an independent fundamental interest.

The complication is that the worldsheet theory necessarily has gapless excitations at $D>2$. 
As a result, unlike in a conventional setting,  the threshold for multiparticle production is the same as for the $2\to 2$ scattering and starts at zero energies.  Also polarization effects coming from loops of virtual particles traveling ``around the world" (also
called winding corrections) are not exponentially suppressed. 

Perhaps, these are the reasons that traditionally \cite{Luscher:1980ac,Luscher:2004ib,Aharony:2010db} the worldsheet spectral data measured on a lattice has been treated quite differently from, for example, pions spectral data. Namely, using the low energy effective string theory (see, e.g., \cite{Dubovsky:2012sh,Aharony:2013ipa} for the introduction) one were to calculate the string spectrum in the $\ell_s/R$ expansion\footnote{Throughout the paper $\ell_s^{-2}\sim \Lambda_{QCD}^2$ is the string tension, and $R$ is a circumference of a circle wrapped by the string, which sets the compactification size of the worldsheet theory.} and compare it with the lattice data.  Unfortunately,
 the $\ell_s/R$ expansion has very poor convergence properties for excited states of a string. As a result, with the existing lattice data the applicability of
this  technique is mostly limited to the ground state.

To get around this problem one switches to a calculational scheme relating finite volume spectrum to scattering amplitudes  \cite{Dubovsky:2013gi,Dubovsky:2014fma}.
In this approach one uses a low energy effective theory to calculate perturbatively the worldsheet $S$-matrix. The transition from the $S$-matrix to the finite volume spectrum is performed non-perturbatively. At the leading order in the derivative
expansion the effective theory is described by the Nambu--Goto action. The corresponding tree level amplitudes are integrable ({\it i.e.}, there is no particle production). This allows to apply  the Thermodynamic Bethe Ansatz (TBA) \cite{Zamolodchikov:1989cf,Dorey:1996re} to exactly reconstruct the corresponding energy spectrum even though the theory is massless.

This approach allowed to identify a massive pseudoscalar resonance (``the worldsheet axion") on a worldsheet of $D=4$ confining strings and led to  the Axionic String Anstaz (ASA) \cite{Dubovsky:2015zey,Dubovsky:2016cog,Dubovsky:2018vde} for the structure of the worldsheet theory both at $D=3$ and $D=4$. To make  further use of lattice data one needs to extend this approach to multiparticle states and find a systematic
way to incorporate higher order non-integrable corrections to the worldsheet scattering. Ideally, one would like to be able to directly reconstruct scattering amplitudes bypassing the effective field theory calculation. Several steps towards achieving these goals were made in  \cite{Dubovsky:2013gi,Dubovsky:2014fma}, but a lot remains to be done.

In the present paper we focus on  winding corrections. Namely, we describe a fully systematic and simple recipe to account for the polarization effects associated with the leading order contribution to  scattering amplitudes.
The recipe applies for states with an arbitrary number of particles and in the presence of arbitrary higher order non-integrable interactions. Note that  the  polarization corrections can be also described as  effects associated with a thermal bath of
temperature $T=R^{-1}$. As a result these are less sensitive to the UV behavior of the theory than the effects associated to the scattering of real particles and accounting for polarization effects related to the leading order interactions is often all one needs
(see section 3.4 of  \cite{Dubovsky:2014fma} for the detailed version of this argument, and sections 4.2, 4.3 for explicit examples illustrating the smallness of subleading winding corrections).

The recipe presented here is based on several recent theoretical developments. First, at the level of scattering amplitudes it has proven very convenient  to think about the worldsheet $S$-matrix in terms of the following ``gravitational dressing" \cite{Dubovsky:2013ira,Conkey:2016qju}. Namely, the worldsheet $S$-matrix can be written in the form
\be
\label{dressed}
S=e^{i\ell_s^2 P_LP_R}S_u\;,
\ee
where $S_u$ is the ``undressed" $S$-matrix, and $P_L$  ($P_R$) is the total momentum of left(right)-moving colliding particles\footnote{Expression (\ref{dressed}) applies when all scattering particles have zero mass, which is the case relevant for the present paper.}. With  the (un)dressing parameter $\ell_s^2$ equal to the inverse string tension, as chosen in (\ref{dressed}), the undressed $S$-matrix $S_u$ is trivial at the leading order in the momenta of colliding particles. 

The representation (\ref{dressed}) is useful for our purposes here because, as proven in \cite{Dubovsky:2017cnj,Dubovsky:2018bmo}, the gravitational dressing (\ref{dressed}) is equivalent
to the $T\bar{T}$ deformation introduced in \cite{Smirnov:2016lqw,Cavaglia:2016oda}, building up on \cite{Zamolodchikov:2004ce}. This equivalence implies that  the finite volume spectra of the worldsheet  and the undressed theories are related by the following ``hydrodynamical" differential equation \cite{Smirnov:2016lqw,Cavaglia:2016oda},
\be
\label{hydro}
\d_{\ell^2}E_n={1\over 2}\l E_n\d_R E_n+{P_n^2\over R} \r\;.
\ee
Here 
\[
P_n={2\pi k_n\over R}
\] 
is the total momentum of a state $n$ and $E_n(\ell^2,R)$ is a family of the corresponding energies labeled by the dressing parameter $\ell^2$. Physical energies of the worldsheet theory are obtained by setting
$\ell^2=\ell_s^2$, while the energies of the undressed theory correspond to $\ell^2=0$. Let us stress that the hydrodynamical equation (\ref{hydro}) provides the exact relation between the spectra of the worldsheet and the undressed theories, accounting both for real scattering and for polarization effects associated with the dressing factor in (\ref{dressed}).

This leads to the following strategy for calculating the finite volume spectrum on the worldsheet. One calculates perturbatively the worldsheet $S$-matrix $S$ and reconstructs the corresponding undressed amplitudes $S_u$ using (\ref{dressed}). Then one calculates the finite volume spectrum $E_n(0,R)$ of $S_u$, using either Asymptotic Bethe Ansatz (ABA) (which is essentially the same as the L\"uscher quantization condition), or some other approximation. In some examples below, where a proper generalization of the L\"uscher equations has not yet been  developed, we use a hybrid of the ABA and of the $\ell_s/R$ expansion for inelastic multiparticle scattering. Finally, one accounts for the leading order scattering and polarization effects by solving (\ref{hydro}) using $E_n(0,R)$ obtained at the previous step as an initial condition.

It is important to stress the following fact. In principle, the above procedure can be implemented in any two dimensional theory. It amounts to reorganizing the perturbative expansion around a $T\bar{T}$ deformed free theory rather than just around a free one, as is usually done. In general, there is no reason to expect that this reorganization of the perturbation theory provides any mileage (however, it would still be a fully systematic procedure, even if unnecessarily complicated).
However, in the case of  the worldsheet theory there is a strong motivation to adopt this procedure.
Namely, a special property of the string worldsheet theory is that the undressed theory in this case is free at the leading order in the derivative expansion---undressing removes all vertices containing one derivative per field!

Indeed, as a consequence of the nonlinearly realized target space Poincar\'e symmetry, 
before dressing all leading order interactions in the worldsheet theory are given by the  Nambu--Goto action. The corresponding tree level amplitudes are reproduced by the expansion of the dressing factor up to the corresponding order in $\ell_s$ \cite{Dubovsky:2012wk}, so that the undressed $S$-matrix may only contain higher order interactions. Alternatively, one can deduce this statement directly at the level of the action from the results of \cite{Caselle:2013dra,Cavaglia:2016oda}. This kind of arguments are explained in detail in \cite{Conkey:2016qju}, where they are proven to provide a powerful tool for multiloop calculations in the worldsheet theory.

In the rest of the paper we illustrate this prescription with several examples. Namely, in section~\ref{D4} we apply it to the $D=4$ Yang-Mills data and reproduce the results of  \cite{Dubovsky:2013gi}. We derive the prediction for the energy spectrum of two particle states following from the minimal Nambu--Goto effective action and then show the effect of the wordsheet axion. In section~\ref{D3} we turn to  $D=3$ Yang-Mills. We first consider energy splittings between two and four particle states previously analyzed 
in \cite{Dubovsky:2014fma}. Then we turn to three particle states, whose energy splittings at the leading order are controlled by the same higher dimensional operator which appears in the two and four particle  sector.
These splittings provide a probe of non-elasticity, which is shown to grow at large collision energy.
 In section~\ref{last} we present our conclusions. In Appendix~\ref{app:TBA} we illustrate the correspondence between the dressed $S$-matrix (\ref{dressed}) and the deformation equation (\ref{hydro}) for the finite volume spectrum in the integrable case, where the spectrum can be found from the TBA equations.
 In Appendix~\ref{PS} we present an efficient way to calculate leading order amplitudes with an arbitrary number of particles corresponding to  any higher dimensional operator in the effective string action. 
\section{$D=4$ Yang--Mills}
\label{D4}
Consider a long confining string stretched in the $X^1$ direction in $D=4$ Yang--Mills theory. It carries two gapless modes corresponding to excitations in the transverse directions $X^i$'s, $i=2,3$. The low energy dynamics of these modes
is governed by the Nambu--Goto action,
\be
\label{NG}
S_{NG}=-{1\over\ell_s^2}\int d^2\sigma\sqrt{-\det\l\eta_{\alpha\beta}+\ell_s^2\d_\alpha X^i\d_\beta X^i\r}+\dots
\ee
where dots stand for higher dimensional operators. A peculiar property of this effective field theory is that unlike for the pion chiral Lagrangian, the first non-trivial counterterm arises only at the two loop order (or equivalently, at the next-to-next-to-leading order in the derivative expansion). Hence both tree level and one loop amplitudes are completely determined by the leading order action (\ref{NG}). Tree level amplitudes following from (\ref{NG}) are purely elastic. The first non-elastic process is one loop $2\to 4$ scattering, and the corresponding amplitude arises at the ${\cal O}(\ell_s^6)$ order. Restricting to the  ${\cal O}(\ell_s^4)$ order one obtains the following elastic two-particle $S$-matrix \cite{Dubovsky:2012sh}
\be
\label{Sls4}
S_{\ell_s^4}=1+i\ell_s^2p_l p_r+\ell_s^4{(p_lp_r)^2\over 2}\l-1\pm {11i\over 6\pi}\r+{\cal O}(\ell_s^6)\;.
\ee
Here $p_l$ and $p_r$ are momenta of the colliding particles. The plus sign in (\ref{Sls4}) describes  scattering in the scalar and pseudoscalar channels w.r.t. the $O(2)$ group of rotations in the transverse plane. The minus sign in
(\ref{Sls4}) corresponds to the spin 2 channel.  Comparing (\ref{Sls4}) to (\ref{dressed}) we find that the undressed $S$-matrix in this case is
\be
\label{Slsu4}
S_{u,\ell_s^4}=1\pm i\ell_s^4 {11\over 12\pi}(p_lp_r)^2+{\cal O}(\ell_s^6)\;.
\ee

\begin{figure}[t!]
  \begin{center}
        \includegraphics[height=6cm]{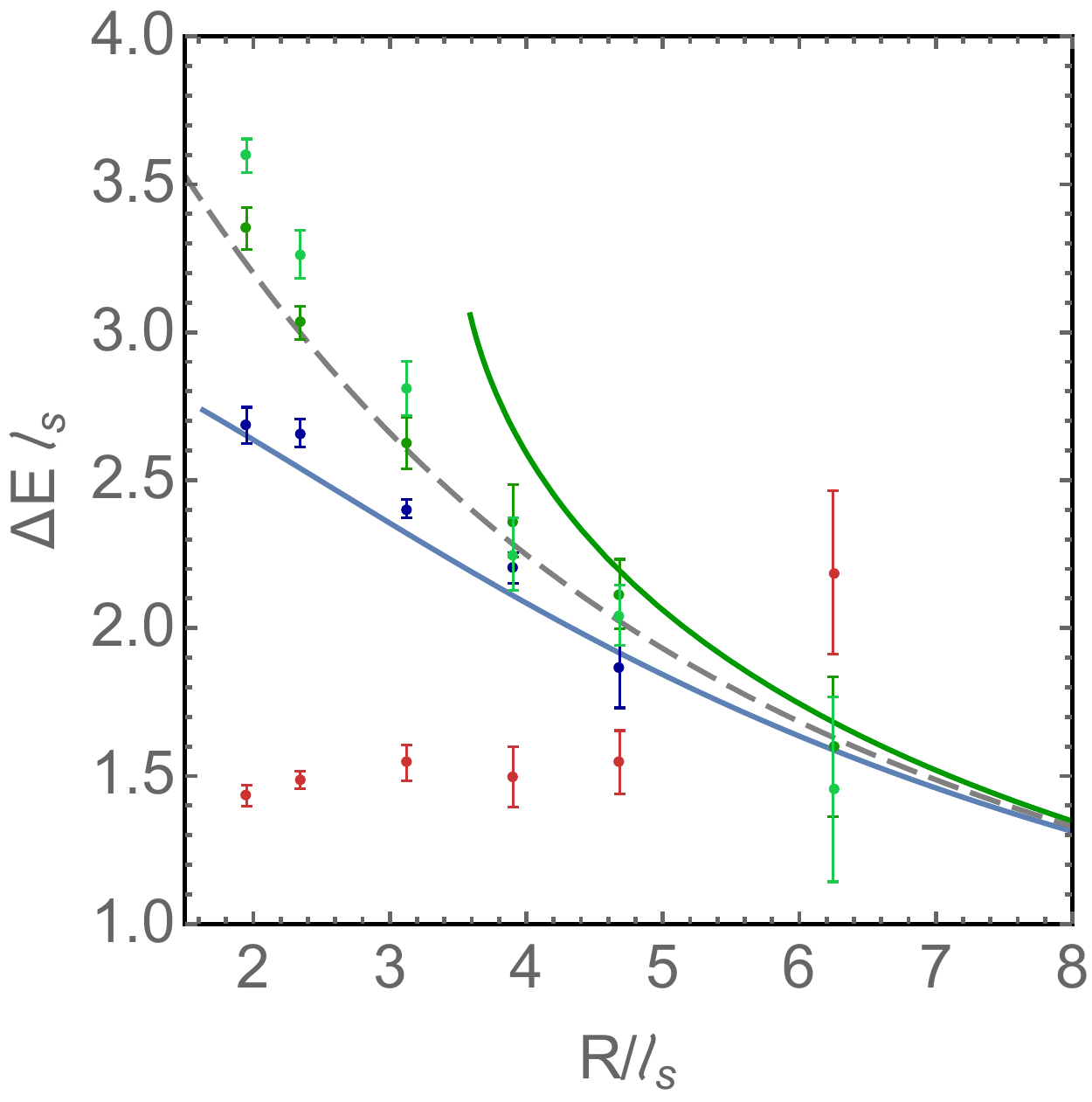} \hspace{2cm}
         \includegraphics[height=6cm]{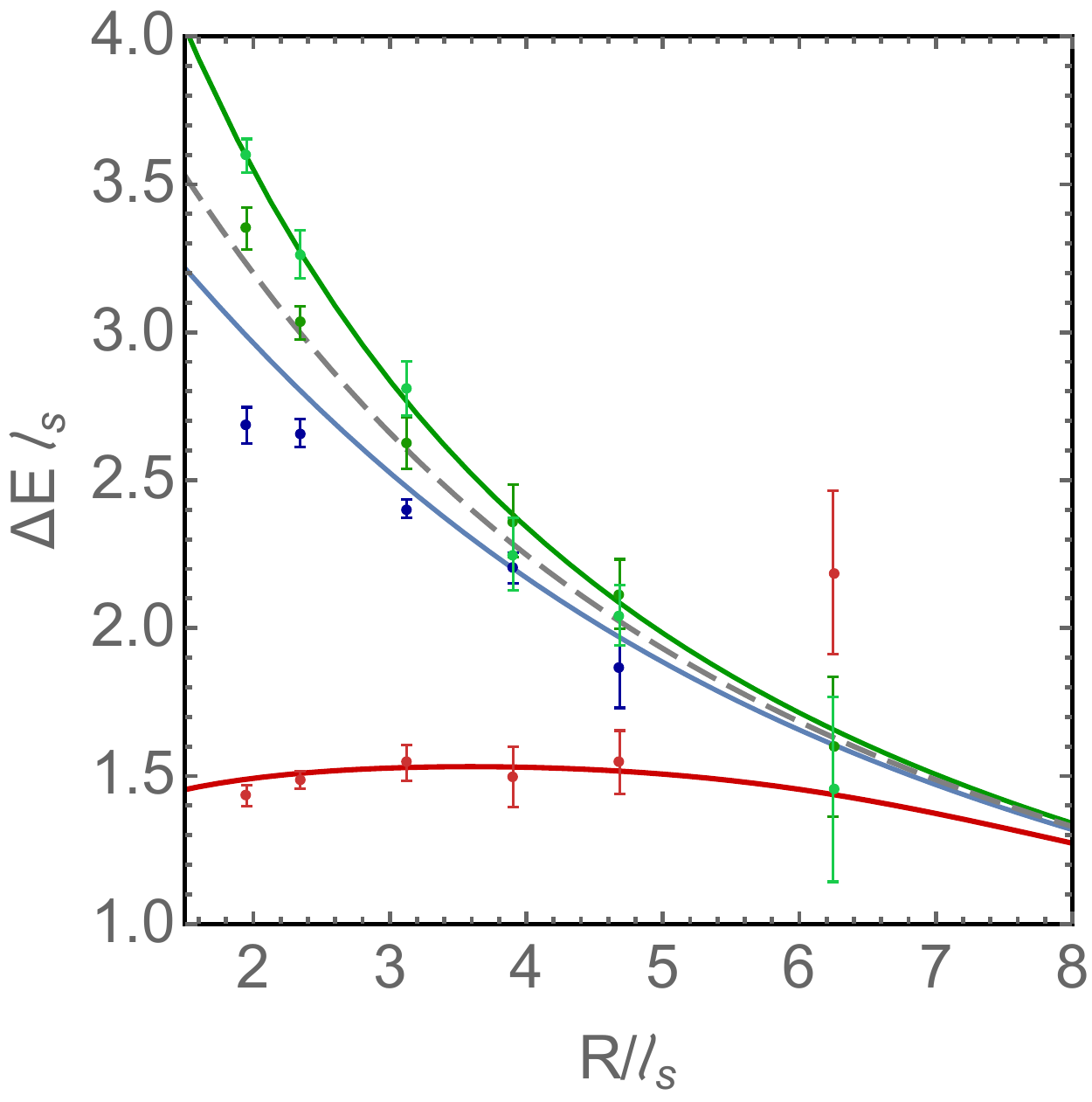} 
           \caption{ The energy gap between  the lowest  two particle excitations and the ground state on the worldsheet. Blue color refers to scalar, red to pseudoscalar and green to spin 2 excitations w.r.t. to the transverse $O(2)$ rotation group. The left panel shows one loop predictions of the minimal Nambu--Goto theory (in this case pseudoscalar and scalar levels are predicted to be degenerate), and the right panel includes the effect of the worldsheet axion. Dashed lines on both panels show the tree level Nambu--Goto prediction (all states are degenerate in this approximation). Lattice data is from \cite{Athenodorou:2010cs}.}
        \label{fig:PS}
    \end{center}
\end{figure}

To illustrate the undressing technique let us calculate now the spectrum
of two particle states with zero total momentum so that 
\[
p_l=p_r\equiv p\;.
\]
Following the recipe outlined in the {\it Introduction} we start with calculating the corresponding spectrum in the undressed theory. In the approximation (\ref{Slsu4}) one can actually completely diagonalize the factorized $S$-matrix for any number of particles by switching to the helicity basis. This allows to write the full set of TBA equations at this order  (for details see
\cite{Dubovsky:2014fma}, where this has been explained directly in the worldsheet theory).
However, one finds that the effect of the phase (\ref{Slsu4}) on the polarization effects is negligibly small in agreement with the general argument about their UV insensitivity . Hence, we will use the ABA approximation which reduces to the following quantization condition (aka the L\"uscher equation),
\be
\label{LuscherPS}
pR+2\delta_{PS}(p)=2\pi n \;,
\ee
where
\be
\label{dPS}
 2\delta_{PS}(p)=\pm\ell_s^4 {11\over 12\pi}p^4
 \ee
is the phase shift corresponding to (\ref{Slsu4}), and  $n$ is a positive integer. For the lowest lying two-particle excitations $n=1$. The PS subscript refers to the fact that this phase shift describes the effect of the Polchinski--Strominger term 
\cite{Polchinski:1991ax} (see \cite{Hellerman:2014cba} for a nice exposition of the PS formalism, and \cite{Dubovsky:2012sh,Dubovsky:2016cog} for the explanation of how 
it is related to the phase shift (\ref{dPS})).
Given the solution $p(R)$ of (\ref{LuscherPS}) the corresponding undressed finite volume energy is given by
\be
\label{Eu}
E_u(R)=2p(R)-{\pi\over 3R}\;,
\ee
where the last term is the Casimir energy of two free massless bosons. The last remaining step is to solve the hydrodynamical equation (\ref{hydro}) 
using  (\ref{Eu}) as the initial condition. For  vanishing total momentum, $P_n=0$, which is the case at hand, the implicit solution to (\ref{hydro}) takes the following simple form
\cite{Smirnov:2016lqw,Cavaglia:2016oda},
\be
\label{impl}
E(R,\ell_s^2)=E_u\l R+{\ell_s^2\over 2}E(R,\ell_s^2)\r\;.
\ee
The corresponding energies are shown in the left panel of Fig.~\ref{fig:PS}. They are in a complete agreement with the spectrum obtained in \cite{Dubovsky:2013gi,Dubovsky:2014fma},
as it should be.
At this order pseudoscalar and scalar states are predicted to be degenerate in the minimal Nambu--Goto theory. Clearly, this expectation is in conflict with the lattice data, which exhibits an anomalously light pseudoscalar state. It is exactly this plot which motivated \cite{Dubovsky:2013gi} to introduce the worldsheet axion.
It is straightforward to reproduce the resulting spectra with the undressing technique. Namely, the axion leads to an additional contribution to the phase shift, which takes the form 
\be
\label{dres}
2\delta_{res}(p)=2\sigma_2\tan^{-1}\l {8Q_a^2\ell_s^4p^6\over m^2-4p^2}\r+\sigma_1{8Q_a^2\ell_s^4p^6\over m^2+4p^2}
\ee
where $Q_a$ is the axion coupling constant (we follow the conventions of \cite{Dubovsky:2015zey}), $m$ is the mass, and $\sigma_1=(-1,1,1)$, $\sigma_2=(0,0,1)$ for scalar, spin 2, and pseudoscalar channels approximately. It is straightforward to incorporate this phase shift into the quantization condition of the undressed theory (\ref{LuscherPS}), and evaluate the corresponding undressed energies, which as before can be dressed using (\ref{impl}). The result is shown in the right panel of Fig.~\ref{fig:PS}. Here we chose the best fit parameters of 
\cite{Dubovsky:2013gi,Dubovsky:2015zey},
\[
m=1.85\ell_s^{-1}\;,\;\;Q_a=0.37\;.
\]
Again, we find  perfect agreement with the results obtained directly in the worldsheet theory.
This time it is slightly more non-trivial, given that the polarization effects due to the axion cannot be incorporated in the TBA so one cannot check directly that they are small.
So the agreement in this case can be considered as a test that they are indeed small, as expected on general grounds.

It is straightforward to extend this analysis to states with non-zero total momentum. The corresponding generalization of the implicit solution 
(\ref{impl}) is given by equation \eqref{implicitE_pnonzero}, although in practice it is simpler to solve equations \eqref{implicitEl_pnonzero} and \eqref{implicitEr_pnonzero}. Instead, let us consider multiparticle states as our second example, using this time the $D=3$
flux tube spectra.

\section{$D=3$ Yang--Mills}
\label{D3}
\begin{figure}[t!]
	\begin{center}
		\includegraphics[height=6cm]{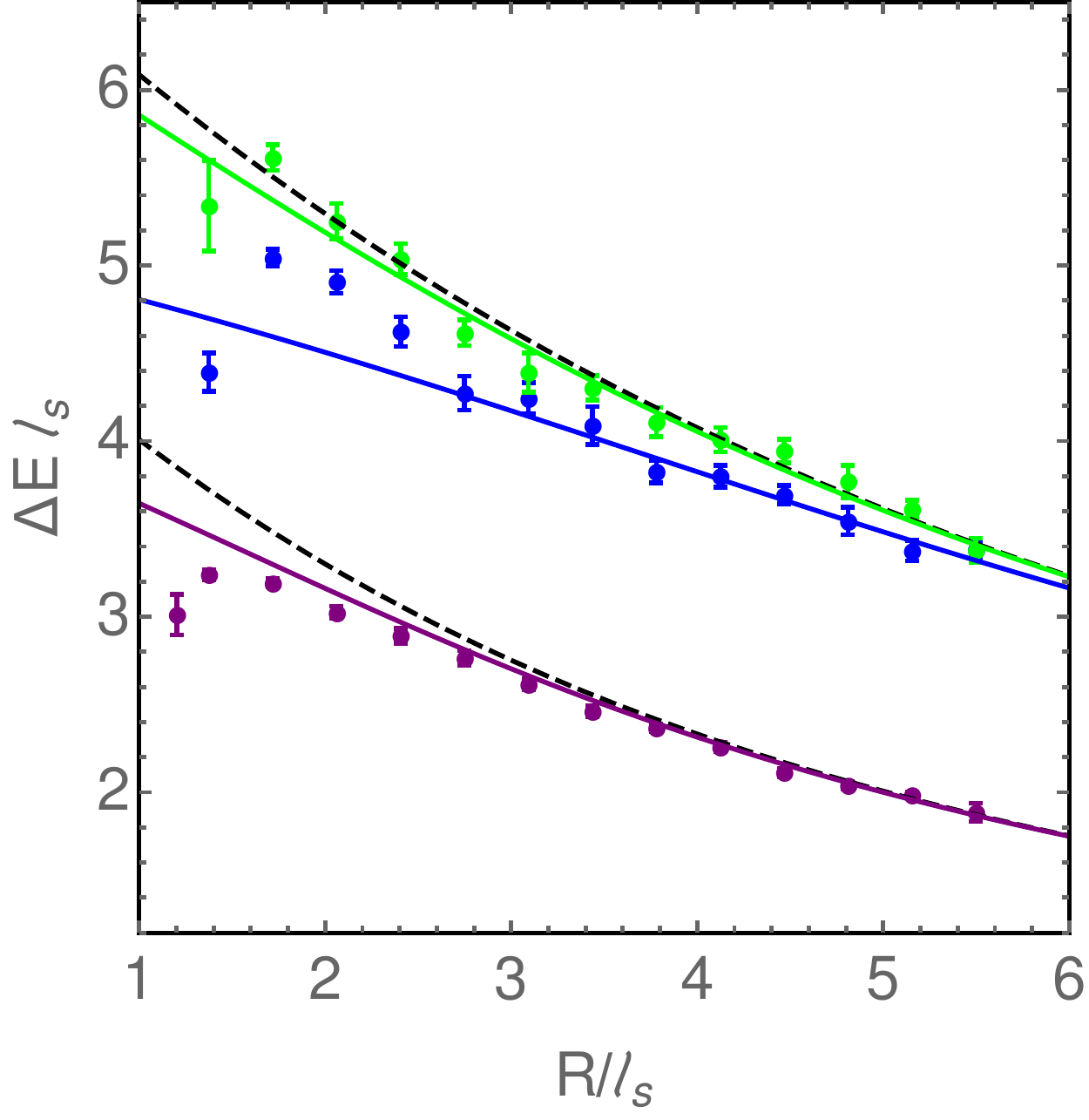} \hspace{2cm}
		\includegraphics[height=6cm]{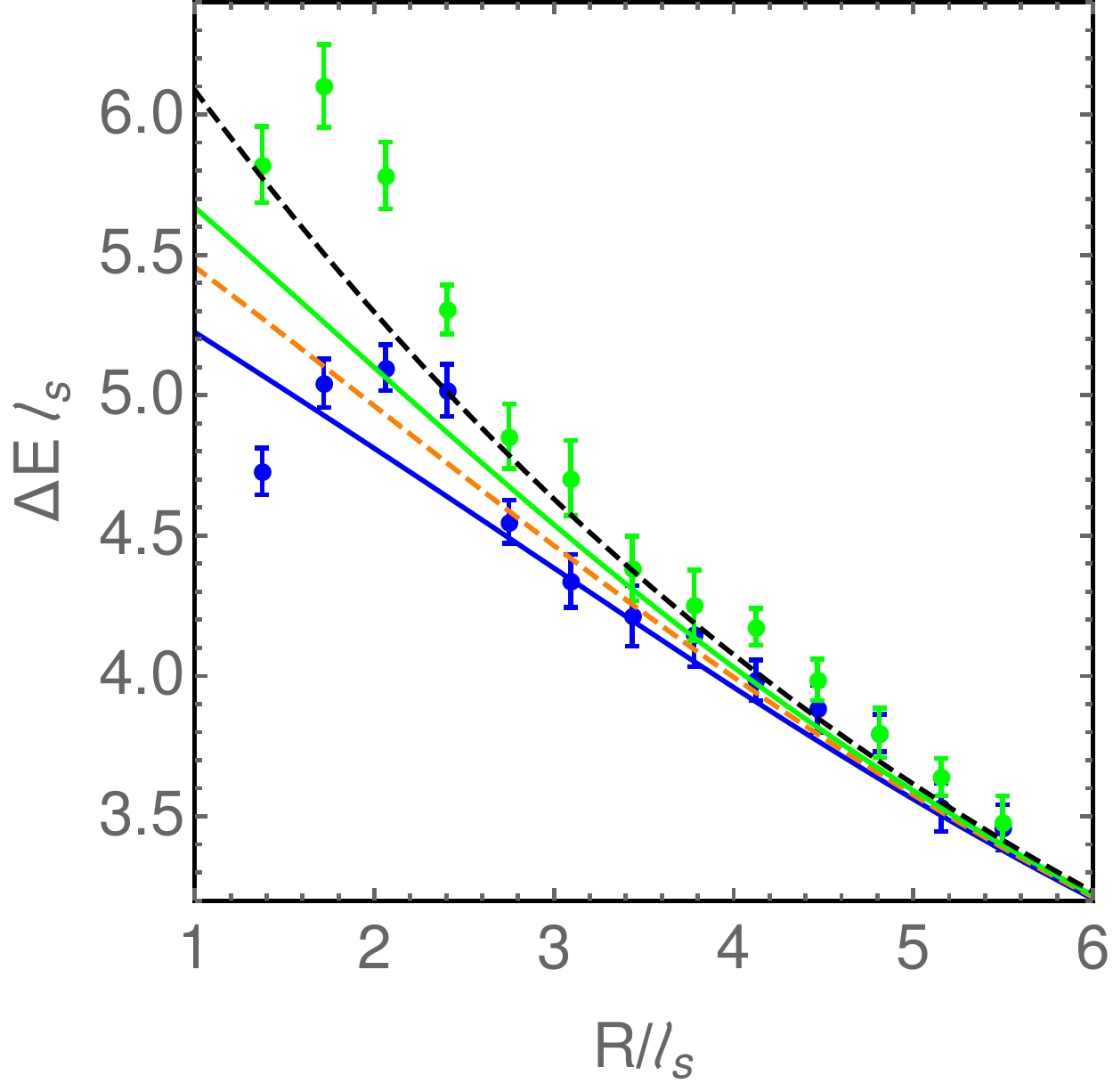} 
		\caption{Lattice data and fits for $\ell_s\Delta E \equiv \ell_s E - R/\ell_s$ as a function of $R/l_s$ for the $N=1,2$ states of the flux-tube. Dashed black lines show the GGRT spectrum. The left panel shows the first and second excited states with two particles (purple and blue markers) and the first excited state with four particles (green markers). The solid colored lines stand for the corresponding theoretical curves. The right panel shows the first two three particle excited states. The orange dashed line results from the dressed ABA calculation. The green and blue lines correspond to including perturbatively the effects of the inelasticity.}
		\label{fig:3Ddata}
	\end{center}
\end{figure}

At $D=3$ the low energy dynamics  is again governed  by the Nambu--Goto action (\ref{NG}), but this time with a single Goldstone field $X$. There is no analog of the PS amplitude in this case because the corresponding one loop contribution vanishes for kinematical reasons. Hence, the theory is integrable at one loop level and the corresponding ${\cal O}(\ell_s^4)$ two particle $S$-matrix is the same as for a dressed massless boson,
\be
S_{\ell_s^4}=e^{i\ell_s^2 p_lp_r}\l1+{\cal O}(\ell_s^6)\r\;.
\ee
The corresponding spectrum (which can be calculated either using the TBA or by solving the hydrodynamical equation (\ref{hydro})) is the same as one obtains by performing the light cone quantization 
in the  sector with winding and takes the form
\begin{equation}
E\left(N,\tilde{N}\right) = \frac{1}{\ell_s}\sqrt{\frac{R^2}{\ell_s^2} +\frac{4\pi^2\ell_s^2(N-\tilde{N})^2}{R^2} + 4\pi\left(N+\tilde{N}-\frac{1}{12}\right)}\;,
\label{ec:GGRT_spectrum}
\end{equation}
where $2\pi N/R$ and $2\pi \tilde{N}/R$ are the total left- and right-moving momenta of the string. 
In Fig.~\ref{fig:3Ddata} we plotted the $SU(6)$ flux tube spectra measured in \cite{Athenodorou:2011rx}\footnote{A more recent $D=3$ data is presented in \cite{Athenodorou:2016kpd}. However, this newer data provides a more detailed measurement of the spectra at shortest length at $R\lesssim 3\ell_s$, but is less accurate at longer $R$. Given that the low energy effective field theory is of little use at those short $R$ for most of the states,  only the older data is presented in Fig.~\ref{fig:3Ddata}.}. The left panel shows exictations with even number of particles corresponding to $N=\tilde{N}=1$ and $N=\tilde{N}=2$ states. The right panel shows three-particle states with 
$N=\tilde N=2$. 

\subsection{Even Parity Sector and a Modified Phase Shift}
Note that at $D=3$ states with even and odd number of particles  have even and odd parity w.r.t. $X\to -X$ reflection symmetry.
Let us start our analysis with the parity even sector. We see that the spectrum (\ref{ec:GGRT_spectrum}) (which is sometimes referred to as the Nambu--Goto spectrum, or 
the Alvarez--Arvis spectrum \cite{Alvarez:1981kc,Arvis:1983fp}, or the Goddard--Goldstone--Rebbi--Thorn (GGRT) spectrum \cite{Goddard:1973qh}) provides a quite good approximation to the data. However, it is clear from this plot that the flux tube spectra are not given just by (\ref{ec:GGRT_spectrum})\footnote{Note, that at $D=3$, unlike at $D=4$, the GGRT spectrum (\ref{ec:GGRT_spectrum}) is in principle compatible with the non-linearly realized target space Poincar\'e symmetry \cite{Dubovsky:2015zey}.}.

The two loop $2\to 2$ amplitude in the Nambu--Goto theory has been calculated in \cite{Conkey:2016qju} and takes the following form,
\be
\label{2loops}
S_{\ell_s^6}=e^{i\ell_s^2 p_lp_r}\l1+i\gamma \ell_s^6 p_l^3p_r^3+{\cal O}(\ell_s^8)\r\;.
\ee
Note that the first non-trivial higher dimensional operator in the worldsheet action is
\be
\label{R2}
S_{{\mathcal R}^2}=\ell_s^2 \int \sqrt{-h}{\mathcal R}^2\;,
\ee
where 
\[
h_{\alpha\beta}=\eta_{\alpha\beta}+\ell_s^2\d_\alpha X\d_\beta X
\]
and ${\mathcal R}$ is the corresponding scalar curvature. This operator contributes at ${\cal O}(\ell_s^6)$ so the value of $\gamma$ in (\ref{2loops}) is not universal.
Using the Nambu--Goto action only and the MS scheme the value of $\gamma$  is 
\be
\label{MS}
\gamma_{MS}={85\over 432\pi^2}\approx {0.787\over (2\pi)^2}\;.
\ee

At this order in the $\ell_s^2$ expansion the scattering is still integrable, so 
 to obtain the undressed spectrum we can use the ABA equations, as was done in the previous section. For 
two and four particle states with vanishing total momentum these take the following form 
 \begin{equation}
R p +  2\delta_u(p) k = 2\pi n
\label{ec:ABAmultipart}
\end{equation}
where $k=1$ for two particle states, and $k=2$ for four particle ones. The lowest  two particle state as well as the lowest  four particle state corresponds to $n=1$, and the first excited two particle state  to $n=2$.
The undressed phase shift $\delta_u$ 
corresponding to \eqref{2loops} is
\begin{equation}
2\delta_u = \gamma\ell_s^6p^6\label{undressed_phase_3d}\;.
\end{equation}
After solving for the momentum from (\ref{ec:ABAmultipart}) one calculates the undressed energy 
\[
E_u=2 k p-{\pi\over 6 R}\;,
\]
and as before uses (\ref{impl}) to find the physical energy.
In the left panel of Fig.~\ref{fig:3Ddata} we plotted the result obtained  using the best fit value  
\be
\label{fit}
\gamma={0.7\pm0.1\over (2\pi)^2}
\ee
found in \cite{Dubovsky:2014fma}. As expected, we again find a perfect agreement with the earlier TBA calculations\footnote{Note that there is a typo in the definition of $\gamma$ in Eq.~(43) in \cite{Dubovsky:2014fma}. One needs to replace $s^3$ there with $p^6$, so that $\gamma_3$ of \cite{Dubovsky:2014fma} is the same as $\gamma$ here. }.  

Note, however, that the corresponding plot presented in Fig.~14 in \cite{Dubovsky:2014fma} looks differently from our Fig.~\ref{fig:3Ddata}a). The reason is that Fig.~14 shows the spectra obtained as a result of using a phenomenological parametrization of $e^{i2\delta_u}$ as a rational CDD factor, which introduces an additional parameter allowing to fit also the points at small $R$. A single parameter fit shown in Fig.~\ref{fig:3Ddata}a) describes reasonably well the states with low and intermediate momenta. However,
the highest momentum states ($n=2$ two-particle states at $R/\ell_s\lesssim 3$) clearly show a tendency to be closer to the unperturbed GGRT spectrum as compared to the expectation based on (\ref{undressed_phase_3d}). This is in broad agreement with the expectation based on the ASA.

For each of the levels in Fig.~\ref{fig:3Ddata} one also observes a dramatic drop off in energy at the very shortest values of $R$,  $R/\ell_s\lesssim 2$. This effect seems likely to be related to the physics associated with the deconfinement phase transition rather than with the worldsheet dynamics. We never use the corresponding points in  our analysis.

 Let us pause  to comment on a previously unnoticed piece of numerology here. Namely, the best fit value (\ref{fit}) is not only of the same order as the Nambu--Goto value (\ref{MS}) (which is expected, and provides a sanity check for (\ref{fit})), but actually agrees with (\ref{MS}) within the error bars. As we said $\gamma$ is not a universal quantity, so  this agreement is most likely a sheer coincidence. The  reason we nevertheless mention it here is that 
 there is yet another somewhat mysterious aspect of the two loop result (\ref{2loops}). Namely, its piece of leading transcedentality ({\it i.e}, the one without $1/\pi^2$ prefactor) exactly matches the expansion of the dressing  exponent, even though
 one may argue that it is also non-universal.  Most likely this can be understood diagrammatically, given that the time delay corresponding to the dressing factor arises as a classical effect in the Nambu--Goto theory. 
 Another related interesting property of the two loop answer (\ref{2loops}) is that the non-universal constant $\gamma$ does not experience any logarithmic running. 
 Still, the agreement between
 (\ref{fit}) and (\ref{MS}) is probably  a coincidence unless one manages to identify an enhanced symmetry at the value of $\gamma$ given by (\ref{MS}).
\subsection{Three Particle States and Inelasticity}
\label{sec:odd}
Turning to the three particle states,   the corresponding undressed ABA equation for the lowest lying three particle states is
\[
pR+2\delta_u(2p,p)=2\pi\;,
\]
where
\[
2\delta_u(p_l,p_r)=\gamma\ell_s^6(p_lp_r)^3\;.
\]
The undressed energy is given by
\[
E_u=4p(R)-{\pi\over 6 R}\;.
\]
The resulting physical energy is shown at  the right panel of Fig~\ref{fig:3Ddata}.
We see that although the $\ell_s^6$ term lifts the degeneracy in the parity even sector, the parity odd states are still degenerate at this order. Indeed,  as long as the phase shift respects  worldsheet parity, the corresponding
ABA energies for these three particle states are going to be degenerate. 

\begin{figure}
	\begin{centering}
		\subfloat[Sextic diagram]{\noindent \centering{}\includegraphics{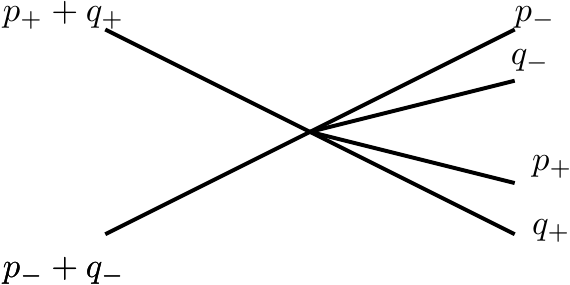}}\hspace{2cm}\subfloat[One quartic Nambu-Goto vertex with one quartic $\sqrt{-h}{\mathcal R}^{2}$ quartic
		vertex (with all permutations of external momenta)]{\noindent \begin{centering}
				\includegraphics{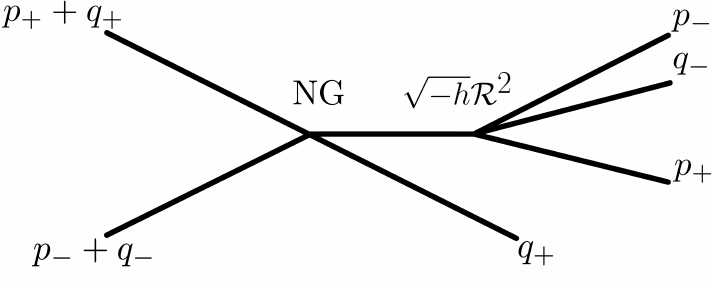}
				\par\end{centering}
			
			\noindent \centering{}}
		\par\end{centering}
	
	\begin{centering}
		\caption{Types of diagrams contributing to 2 to 4 particle amplitude at order $l_{s}^{8}$ \label{fig:Diagrams}}
		\par\end{centering}
\end{figure}

There is however an irreducible splitting among three particle states associated to $\gamma\neq 0$, which arises at order ${\cal O}(\ell_s^8)$ in the scattering amplitude. Namely, the higher-dimensional operator (\ref{R2}) expanded to ${\cal O}(\ell_s^8)$
reads,
\be
\label{sextic}
S_{{\mathcal R}^2}=\int  \frac{\ell_s^6}{4}\left(\partial^2_+X\partial^2_-X\right)^2 + \frac{7\ell_s^8}{8} \left(\partial^2_+X\partial^2_-X\right)^2\partial_+ X\partial_- X
\ee
\noindent where $\partial_\pm \equiv \partial_0-\partial_1$. We find from here that the ${\cal O}(\ell_s^6)$ correction (\ref{undressed_phase_3d}) to the phase shift, implies the presence of proper  ${\cal O}(\ell_s^8)$ six-particle scattering. The corresponding amplitude is given by the sum of the sextic vertex in (\ref{sextic}) and of the tree level diagrams with one  ${\cal O}(\ell_s^6)$ quartic vertex  from  (\ref{undressed_phase_3d}) and one ${\cal O}(\ell_s^2)$ quartic vertex coming from the Nambu--Goto part of the action, see Fig.~\ref{fig:Diagrams}.
A brute force calculation gives the following result for this amplitude

\be
\mathcal{M}_{6}
= -\frac{6\gamma \ell_s^8}{2^7} \left(p_{+}+q_{+}\right)\left(p_{-}+q_{-}\right)p_{+}q_{+}p_{-}q_{-}\left(p_{+}q_{+}+p_{+}^{2}+q_{+}^{2}\right)\left(p_{-}q_{-}+p_{-}^{2}+q_{-}^{2}\right)\label{eq:M6}\;.
\ee
%
Here the overall prefactor is fixed by calculating the four-particle  ${\cal O}(\ell_s^6)$ amplitude associated to the quartic vertex in (\ref{sextic}) and matching it to (\ref{undressed_phase_3d}).

Unfortunately, a three-particle generalization of the ABA quantization condition which would allow to incorporate  amplitude (\ref{eq:M6}) has not been developed yet. Hence, as an estimate, we resort to a mixture of the ABA technique and  $\ell_s/R$ perturbation theory.
Namely, we will calculate the finite volume spectrum of the undressed theory in two steps. First we use ABA to calculate the finite volume spectrum corresponding to the ${\cal O}(\ell_s^6)$ two particle phase shift (this is the calculation we already did). Then we will estimate the splitting between three-particle states by treating (\ref{eq:M6}) perturbatively at the leading order. Finally, as before, we will dress the result.

To implement this strategy it is convenient to start with an action of the undressed theory. It is straightforward to check that the full undressed amplitude at the order we are working (namely, the ${\cal O}(\ell_s^6)$ phase shift  (\ref{undressed_phase_3d})
and the ${\cal O}(\ell_s^8 )$ six particle amplitude (\ref{eq:M6})) follows from the tree level action of the form 

\begin{equation}
S_{u8} = \int\l \frac{1}{2}\partial_+X\partial_-X + \frac{\gamma\ell_s^6}{2^7} \left(\partial^2_+X\partial^2_-X\right)^2 + \frac{3\gamma\ell_s^8}{2^8} \left(\partial^2_+X\partial^2_-X\right)^2\partial_+ X\partial_- X\r\;.
\label{3Dundlag}
\end{equation}
In Appendix~\ref{PS} we present an efficient technique \cite{Dubovsky:2016cog} based on the PS formalism which allows to obtain this action as well as the leading multiparticle amplitudes to all orders in the number of particles 
bypassing the somewhat tedious diagrammatic calculation which led us here.

 To calculate the leading order effect due to the sextic term in \eqref{3Dundlag}, note that it translates in the following term in the interaction Hamiltonian of the compactified theory 
 
\begin{equation}\label{Hint}
{H}_{6} =  -\frac{3\gamma\ell_s^8}{2^8}\int_0^R d\sigma (\partial^2_+ X)^2 (\partial^2_- X)^2\partial_+ X\partial_- X\;.
\end{equation}

At the first order in perturbation theory we need to calculate the matrix elements of this Hamiltonian among three particle finite volume states obtained previously. 
At the moment we don't know how to perform this step rigorously. As an estimate we resort to the following prescription. 
We assume that (\ref{Hint}) is normal ordered which is equivalent to neglecting
winding corrections associated with this interaction. Then we calculate matrix elements of (\ref{Hint}) in a free theory on a circle of size $\bar R$ (in general different from $R$) and without assuming any relation between the
particle momenta and the size of a circle.
The result is 
\be
\label{H6m}
\bra{p,p,-2p} H_6\ket{-p,-p,2p}_{ABA}= 
- \frac{108\gamma\ell_s^8 p^7}{\bar{R}^2}\;.
\ee  
The ABA subscript is a reminder that this is a matrix element  between the finite volume ABA states.  Now we set  the momenta $p$ to  the ABA values obtained previously, and take  the circle size $\bar{R}$ 
to be given by
\begin{equation}
\bar{R} = R + \frac{d \delta(2p,p)}{dp}\;,
\end{equation}
where  $\delta_0$ is the phase shift of an unperturbed theory ({\it i.e.}, in the absence of (\ref{Hint})).
As a consequence of the dressing formula (\ref{impl}) this prescription is equivalent to approximating the unperturbed phase shift by its Taylor expansion in the vicinity of the ABA solution
\[
2\delta(s)\approx 2\delta(s_0)+2\delta'(s_0)(s-s_0)\;.
\]

We see that the diagonal matrix elements of (\ref{Hint}) vanish, so that  (\ref{H6m}) translates in the following splitting  for the three particle states 
\begin{equation}
\Delta E = \pm \frac{108\gamma\ell_s^8 p^7}{\bar{R}^2}\;,
\end{equation}
around the ABA energies obtained previously. The resulting spectrum is presented in Fig~\ref{fig:3Ddata}b). As before we observe that the splitting is of the right order of magnitude at large and intermediate values of $R$. Just as for even parity states the effective field theory breaks down at the highest momenta, corresponding to $R\lesssim 3\ell_s$.
However, somewhat surprisingly, the overall trend in the odd sector is different. In the even sector the highest momenta states demonstrate the tendency to come closer to the GGRT spectrum. Instead, the splitting in the odd sector
grows with energy and at $R\lesssim 3\ell_s$ becomes even larger than the splitting between four and two particle states in Fig~\ref{fig:3Ddata}a). This is especially surprising, given that the momenta of three particles states are somewhat softer than 
the momenta of $N=2$ two particle states, so {\it a priori} one might expect three particle states to show smaller deviations from the GGRT predictions.

Another interesting aspect of three particle states is that their splitting is necessarily related to the proper six particle interaction. By crossing symmetry the corresponding amplitude $\mathcal{M}_{3\rightarrow 3}$ is equal to the $2\to 4$ amplitude 
$\mathcal{M}_{2\rightarrow 4}$. Hence, an anomalously large splitting between three particle states is a smoking gun of a growing inelasticity at high momenta. To quantify this observation it is instructive to bypass the effective field theory calculation and to use the three particle energy splitting to extract $\mathcal{M}_{3\rightarrow 3}$ directly from the lattice data. Namely, we write
\be
\label{matrix}
\bra{p,p,-2p} H_6\ket{-p,-p,2p}_{ABA}=\Delta E\;,
\ee
where, as before, $\Delta E$ is {\it a half} of the splitting between the two three particle levels.
Then the scattering amplitude is estimated from the matrix element in (\ref{matrix}) using the relation
\be
\label{MRbar}
\mathcal{M}_{3\rightarrow 3}(p,p,-2p\to 2p,-p-p)=-{\bar R^2\over (2\pi)^3}\bra{p,p,-2p} H_6\ket{-p,-p,2p}_{ABA} \;.
\ee
 
\begin{figure}[t!]
	\begin{center}
		\includegraphics[height=6cm]{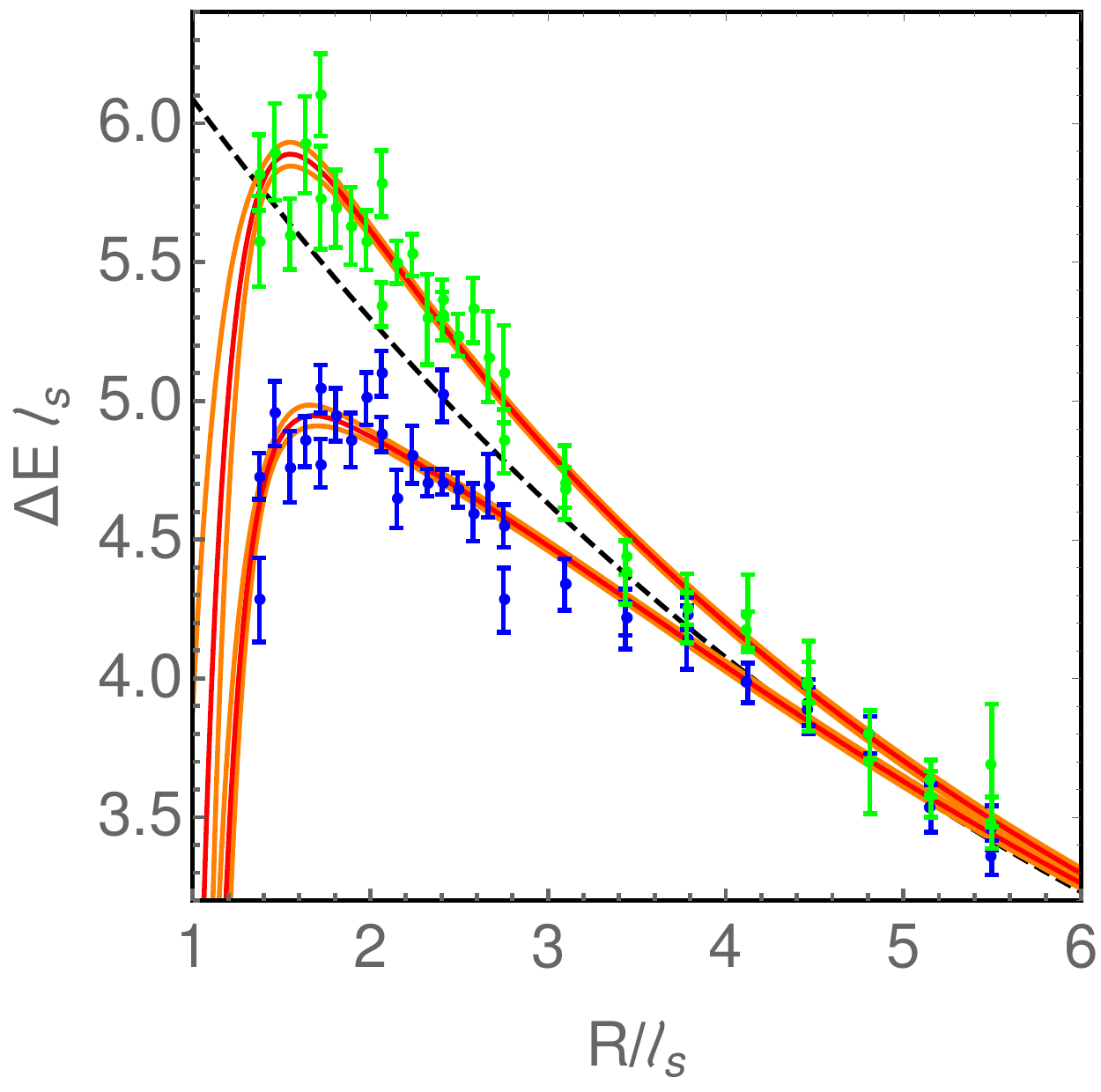} \hspace{2cm}
		\includegraphics[height=6cm]{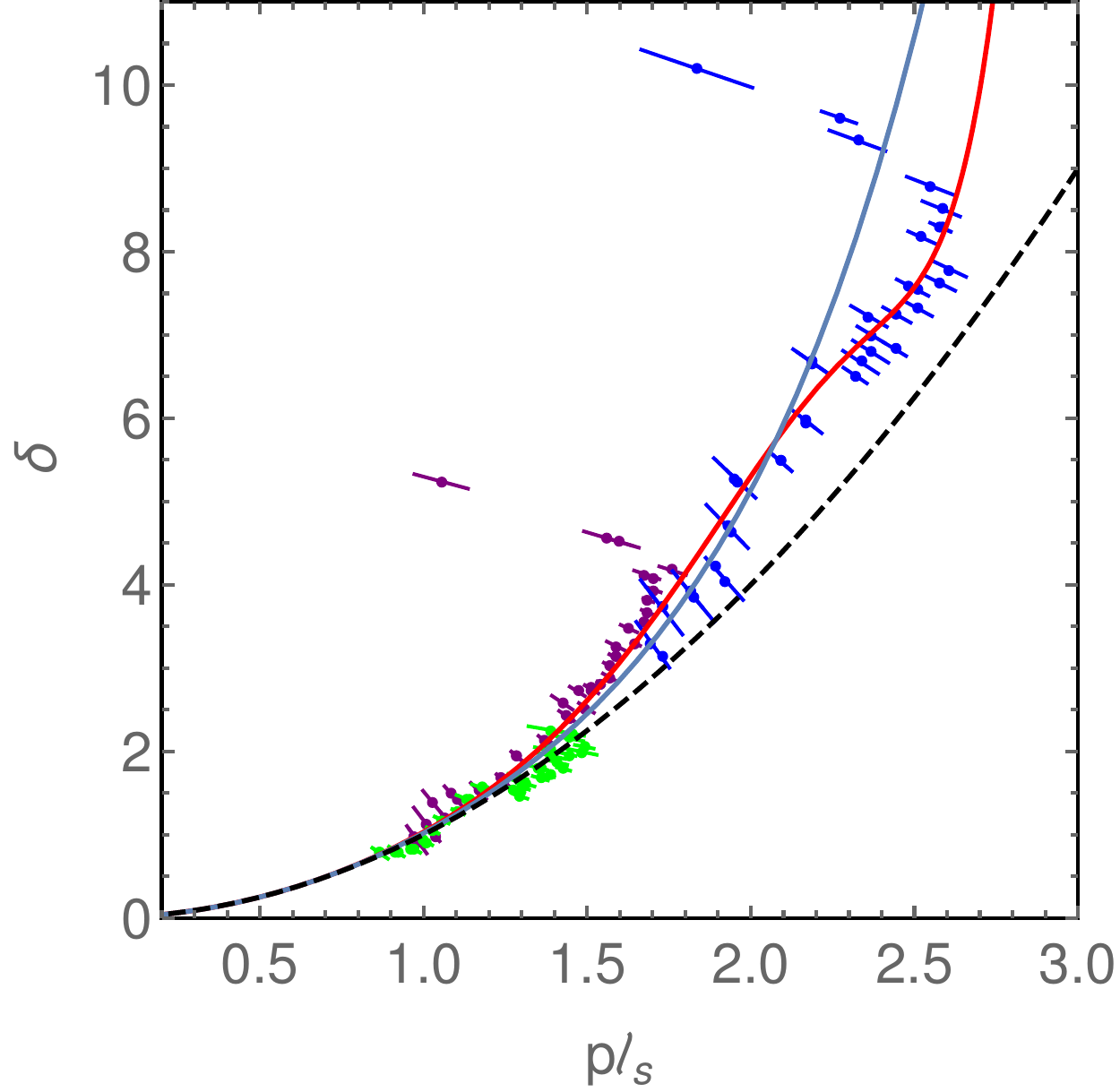} 
		\caption{Input data used for determination of the contact three particle scattering amplitude $\mathcal{M}_{3\rightarrow 3}$.
		a) Curves obtained from the best fit values to the coefficients in (\ref{fitEform}) (red) and $1\sigma$ variations in best fit parameters (orange). b) Dressed phase shift as a function of momentum. The dots represent lattice data points for first and second excited two particle states (purple and blue dots respectively) and first four particle state (green dots). The black curve corresponds to the GGRT phase-shift and the blue curve to the GGRT phase shift with the $\ell_s^6\gamma p^6$ correction. The red curve corresponds to the best fit parameters for the parametrization given by \eqref{fitpsform}.}
		\label{fig:fitstoe}
	\end{center}
\end{figure}

 To perform this extraction in practice it is convenient to parametrize the observed lattice data by a smooth curve. This is just a matter of technical convenience.
 We take the following ansatz for the fitting curve

\begin{equation}
E^{(\pm)}(R) = E_{GGRT}(R) + \displaystyle\sum_{n=2}^{5} a^{(\pm)}_n R^{-n} 
\label{fitEform}
\end{equation}    

 where $(\pm)$ label the two three-particle states and $E_{GGRT}(R)$ stands for the GGRT energy. This ansatz is not based on a theoretical expectation of how the spectrum should depend on $R$. Rather, it just provides a smooth parametrization of a curve that reasonably approximates the data.
For this analysis it is natural to include also the more recent data  from  \cite{Athenodorou:2016kpd}, which is more accurate at the shortest $R$.
In Fig.~\ref{fig:fitstoe}a) we present the resulting smooth curve (red curves) as well as all the data used to perform the fit.
The orange curves indicate the $1\sigma$ uncertainty in the fit.   

In order to obtain the two particle amplitude we also need to extract the undressed phase shift from the data to determine ${\bar R}$ in (\ref{MRbar}). This can be done from the ABA equation (\ref{ec:ABAmultipart}). The resulting physical (not undressed) phase shift is shown 
in Fig.~\ref{fig:fitstoe}b). Here we use lattice data from both \cite{Athenodorou:2011rx} and \cite{Athenodorou:2016kpd}. The latter (newer) data is more accurate at short $R$ corresponding to higher momenta, while the former (older) data 
has smaller error bars at low momenta.
To obtain a smooth approximation we use the following parametrization for the phase shift\begin{equation}
2\delta(p,p) = \ell_s^2 p^2 +\ell_s^6 \gamma p^6 +\sum_{n=4}^{6} b_n p^{2n}\;.
\label{fitpsform}
\end{equation}
\begin{figure}
	\centering
	\includegraphics[height=6cm]{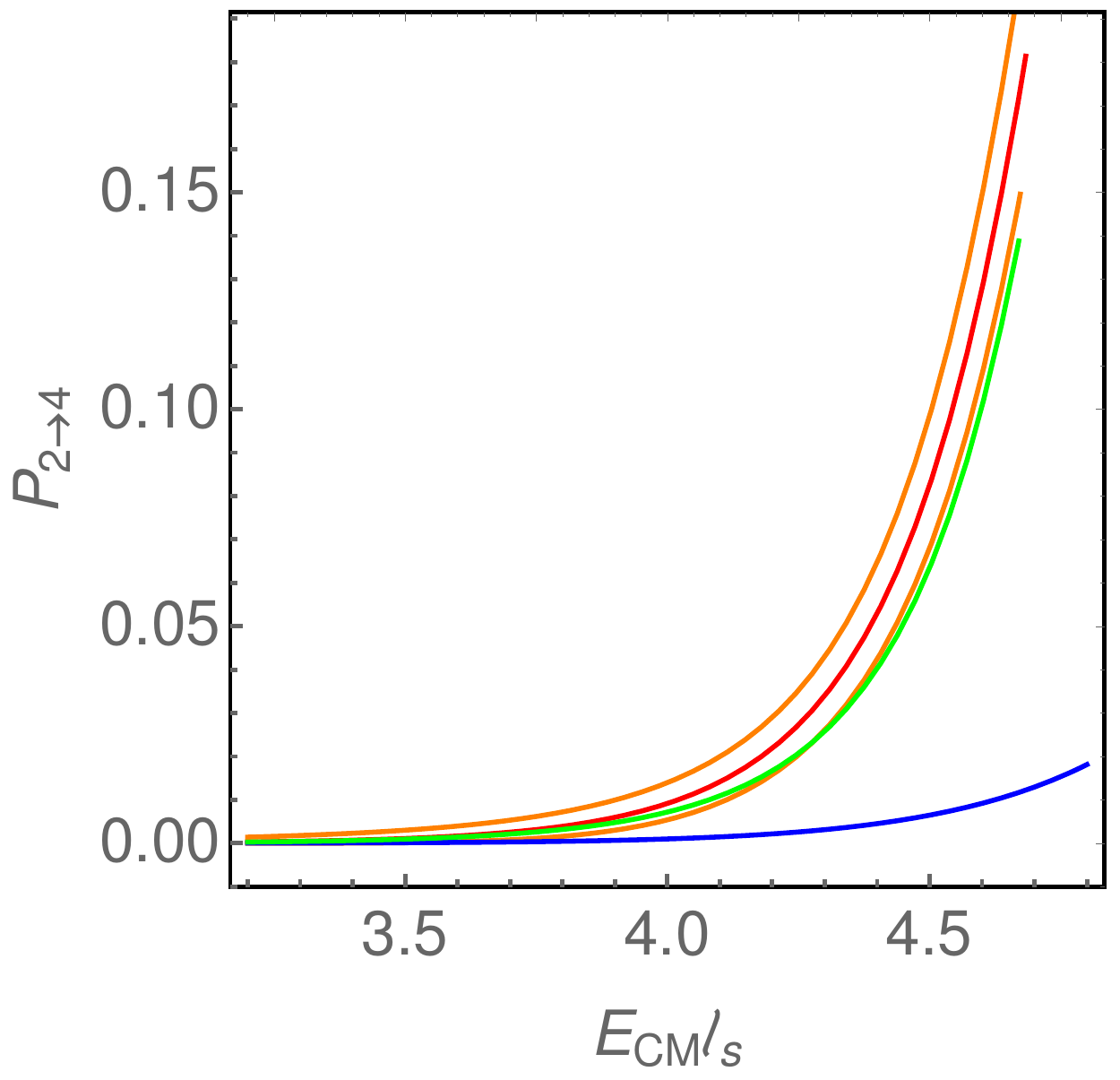}
	\caption{Particle production probability as a function of collision energy. The blue line gives the result from using the undresed lagrangian \eqref{3Dundlag}. The green line is the result extracted from the data using the theoretical $p^6$ phase-shift but fitting the data for the energies. The red curve is obtained by fitting both the phase shift and energies from the data. The orange curves account for fitting uncertainties around this last curve.}
	\label{fig:prob}
\end{figure}
When performing  the fit for $b_n$'s , we exclude the data points at the shortest $R$, $R/\ell_s\lesssim 2$, where the interpretation of the data in terms of the phase shift is clearly non-adequate.
With the phase shift as a function of $p$ and the undressed spectra as a function of $R$ we  use \eqref{fitEform} to obtain $\mathcal{M}_{3\to 3}$, which is also equal to $\mathcal{M}_{2\to 4}$ . 
Once $\mathcal{M}_{2\rightarrow4}$ is known, we calculate the probability for the $2\rightarrow4$ particle production process from an initial state of the form $\left\{2p,-2p\right\}$ 
\begin{equation}
P_{2\rightarrow4}(p) =  \displaystyle\int_{-\infty}^{+\infty}\prod_{i=1,4} dp_i\frac{1}{4!} \frac{(2\pi)^2}{2}\vert\mathcal{M}_{2\rightarrow4}\vert^2\delta\left(\sum_i|p_i|-4p\right)\delta\left(\sum_i p_i\right)\;.
\end{equation}
 Since we only know $\mathcal{M}_{2\rightarrow4}$ for a specific kinematic regime we approximate it by a constant when performing  the phase space integral. This results in the following estimate for the inelasticity
\begin{equation}
P_{2\rightarrow4}(p)\approx  \frac{20(2\pi)^2}{6!2}\left(\frac{\bar{R}(p)\delta E(R(p))}{2(2\pi)^3}\right)^2p^2.
\end{equation}
 This probability is shown by the red line in figure \ref{fig:prob}.
 The orange lines indicate the uncertainty  in the energy fit.  The green line gives the result of using the low energy undressed phase shift $\ell_s^6\gamma p_l^3p_r^3$ when calculating $\bar{R}$ instead of fitting the data.
 The blue line corresponds to using the undressed action \eqref{3Dundlag} to obtain the $3\to 3$  amplitude.   We see that it vastly underestimates the particle production rate.

\section{Future Directions}
\label{last}
To summarize, we see that the $T\bar T$ deformation (equivalently,  gravitational dressing) provides an efficient technique for analyzing the flux tube spectra measured on a lattice. 
It gives a straightforward recipe to account for the leading order polarization effects. 
As should be clear from section~\ref{sec:odd} in order to make full use of the lattice data one needs now to develop generalizations of the ABA equations (L\"uscher formulas)
for multiparticle scattering. This is especially important given the importance of inelastic processes in the high energy scattering on the worldsheet as suggested by theoretical arguments \cite{Dubovsky:2018vde} as well as by the analysis of odd $D=3$ states presented here.

In addition, note that here we simply used gravitational dressing as a convenient technical tool. However, given the expected closed connection between the high energy worldsheet dynamics 
and perturbative QCD \cite{Dubovsky:2018vde} one may wonder whether there is an underlying  gauge theory interpretation of the dressing. The natural answer seems to be that  gravitational dressing is closely related to soft gluon factorization. In this respect it is interesting to note that both phenomena are most easily derived via a field redefinition involving a seminfinite Wilson line. In the $T\bar T$ case it is a gravitational Wilson line in the Jackiw--Teitelboim gravity \cite{Dubovsky:2017cnj,Dubovsky:2018bmo}. In the
soft gluon case it is a conventional gauge theory Wilson line (see, e.g., \cite{Becher:2014oda}). It will be interesting to make this connection precise.
\section*{ Acknowledgements} 
 We thank Andreas Athenodorou,  Raphael Flauger, Victor Gorbenko, Zohar Komargodski, Riccardo Rattazzi and Mike Teper for fruitful discussions. This work is supported in part by the NSF CAREER award PHY- 1352119.

\appendix
\section{Dressing an integrable theory}
\label{app:TBA}

To illustrate how the dressing equation  \eqref{hydro} comes about we present here its derivation, following \cite{Cavaglia:2016oda}, for an integrable theory from 
 the TBA equations.

 Restricting for simplicity to the  single flavor case, the TBA momenta $p_{li}$ and $p_{ri}$ are defined by the conditions (see, for instance, \cite{Dubovsky:2014fma})

\begin{align}
f_l\left(ip_{li}\right) = 2\pi i n_{li}\\
f_r\left(-ip_{ri}\right) = -2\pi i n_{ri}\;,
\end{align}
\noindent where $n_{l(r)i}$ are positive integers and the pseudoenergies $f_{l(r)}$ are determined by the integral equations

\begin{align}
\label{fl}
f_l(q) &= q R + i\sum_i2\delta(q,-ip_{ri}) + \frac{1}{2\pi}\displaystyle\int_0^\infty dq'\frac{d2\delta(q,q')}{dq'}\ln \left(1-e^{-f_r(q')}\right)\\
\label{fr}
f_r(q) &= q R - i \sum_i2\delta(q,ip_{li}) + \frac{1}{2\pi}\displaystyle\int_0^\infty dq'\frac{d2\delta(q,q')}{dq'}\ln \left(1-e^{-f_l(q')}\right)
\end{align}

 To proceed it is convenient to rewrite the r.h.s. of (\ref{fl}), (\ref{fr}) as
 \begin{equation}
f_{l(r)}(q) = Rq + \displaystyle\int_{\mathcal{C}_{l(r)}}dq'\partial_{q'}2\delta\left(q,q'\right)\ln\left(1 -\exp^{-f_{r(l)}(q')}\right)
\label{ec:TBA_pseudoe_contour}
\end{equation}
where the integration contours ${\mathcal{C}_{l(r)}}$ are the joints of an integral over real positive $q$'s and of small contours encircling  the logarithmic singularities
at $f_l\left(ip_{li}\right) = i2\pi n_{li}$ and $f_r\left(-ip_{ri}\right) = -i2\pi n_{ri}$ (with opposite orientations for ${\mathcal{C}_{l}}$ and ${\mathcal{C}_{r}}$).
Similarly, the resulting energies can be written as 
\begin{equation}
\Delta E = E_l + E_r
\label{ec:TBA_E_Erl}
\end{equation}
 with
\begin{equation}
E_{l(r)} = \displaystyle\int_{\mathcal{C}_{l(r)}}dq\ln\left(1 -\exp^{-f_{l(r)}}\right)\;.
\label{ec:TBA_Erl}
\end{equation}

Finally, we will also need the expression for the total momentum $P$. For a system of free particles  $p_i$'s are the physical momenta and the total momentum is
\begin{equation}
P = \frac{2\pi}{R}\sum_i\l n_{li} -n_{ri}\r
\label{ec:totalmom_free}
\end{equation}
The total momentum  in any theory is  quantized in integer multiples of $2\pi/R$. Hence, 
if an interacting   integrable theory can be obtained as a continuous deformation of the free one,  (\ref{ec:totalmom_free})  holds also in the interacting case.

Note now that
 \[
\displaystyle\int_{\mathcal{C}_{l(r)}}dq\partial_qf_{l/r}\ln\left(1 -\exp^{-f_{l/r}}\right) = RE_{l(r)} + \displaystyle\int_{\mathcal{C}_{l}}dq\displaystyle\int_{\mathcal{C}_{r}}dq'2\delta(q,q')\ln\left(1 -\exp^{-f_{l}(q)}\right)\ln\left(1 -\exp^{-f_{r}(q')}\right)
\]
 where we first made use of the expression (\ref{ec:TBA_pseudoe_contour}) for $f_{l(r)}$ and then by (\ref{ec:TBA_Erl}).
As a result,
\begin{equation}
E_l-E_r = \frac{2\pi}{R}\sum_i\l n_{li} -n_{ri}\r + \frac{1}{R}\displaystyle\int_{{\rm I\!R}^+}dq\partial_qf_{l}\ln\left(1 -\exp^{-f_{l}}\right) -  \frac{1}{R}\displaystyle\int_{{\rm I\!R}^+}dq\partial_qf_{r}\ln\left(1 -\exp^{-f_{r}}\right) 
\end{equation}
where we made use of
\begin{equation}
\label{ElmEr}
\displaystyle\int_{\mathcal{C}_{l(r)}}dq\partial_qf_{l(r)}\ln\left(1 -\exp^{-f_{l(r)}}\right) = \displaystyle\int_{{\rm I\!R}^+}dq\partial_qf_{l(r)}\ln\left(1 -\exp^{-f_{l(r)}}\right) + 2\pi\sum_{i}n_{l(r)i}\;,
\end{equation} 
which follows from the definition of the contours ${\mathcal{C}_{l(r)}}$.
 In general, for parity odd states $f_l\neq f_r$, and so {\it a priori} one would not expect the last two terms in the r.h.s. of (\ref{ElmEr})  to cancel. However, they  can be rewritten as contour integrals on the $f$ plane 
\begin{equation}
\left\{\int_{f_l\left({\rm I\!R}^+\right)} -\int_{f_r\left({\rm I\!R}^+\right)}\right\}df\ln\left(1- \exp^{-f}\right)
\end{equation}
and so they cancel whenever the contours $f_l\left({\rm I\!R}^+\right)$ and $f_r\left({\rm I\!R}^+\right)$ may be deformed into each other. 
This is true in the free theory case, so assuming again that an interacting theory can be smoothly deformed to the free one we find that they cancel, implying 
\begin{equation}
P = E_l- E_r\;.
\label{ec:TBA_P_pseudoE}
\end{equation}
 
 Gravitationally dressing an integrable theory amounts to modifying the two particle phase shift by 
\begin{equation}
2\delta(q,q') \rightarrow 2\delta_\lambda(q,q') + \ell_s^2 qq'.
\end{equation}
 Plugging in the dressed phase shift into the defining equation for the  $\ell_s^2$-dependent pseudoenergies we get
\begin{equation}
\label{dressedf}
f_{l(r)}(q,\ell_s^2) = q(R + \ell_s^2 E_{r(l)}) + \displaystyle\int_{\mathcal{C}_{l(r)}}dq'\partial_{q'}2\delta\left(q,q'\right)\ln\left(1 -\exp^{-f_{r(l)}(q')}\right).
\end{equation}
 Let us first consider the case when $P=0$. Then from (\ref{ec:TBA_P_pseudoE})
 we find that   $E_r =E_l=E/2$ and (\ref{dressedf}) implies 
\begin{equation}
E(\ell_s^2,R) = E\left(0,R+ \frac{\ell_s^2}{2} E\left(\ell_s^2,R\right)\right),
\end{equation}
which is equivalent to the hydrodynamical  dressing equation for $P=0$.

To take care of the $P\neq0$ case it is convenient to define
\begin{align}
\bar{f}_l(q,\ell_s^2)\equiv f_l(a q,\ell_s^2)\\
\bar{f}_r(q,\ell_s^2)\equiv f_r(a^{-1}q,\ell_s^2)
\end{align}
where
\begin{equation}
a(\ell_s^2,R) \equiv \sqrt{\frac{R + \ell_s^2 E_l(\ell_s^2,R)}{R + \ell_s^2 E_r(\ell_s^2,R)}}.
\end{equation}
Using that $\delta(aq,q') = \delta(q,aq')$ due to Lorentz invariance, we see from equations \eqref{dressedf} that $\bar{f}_l(q,\ell_s^2)$ and $\bar{f}_r(q,\ell_s^2)$ satisfy the same equations as the undressed case but with $R$ replaced by $\bar{R}$, which is given by
\begin{equation}
\bar{R} \equiv \sqrt{(R + \ell_s^2 E_l(\ell_s^2,R))(R + \ell_s^2 E_r(\ell_s^2,R))}.
\end{equation}
Therefore, using definitions \eqref{ec:TBA_Erl}, we see that the dressed $E_l$ and $E_r$ must satisfy the equations
\begin{align}
E_l(\ell_s^2,R) &= E_l(0,\bar{R}) a(\ell_s^2,R)\label{implicitEl_pnonzero}\\
E_r(\ell_s^2,R) &= \frac{E_r(0,\bar{R})}{a(\ell_s^2,R)}
\label{implicitEr_pnonzero}
\end{align}
Using equations \eqref{ec:TBA_E_Erl} and \eqref{ec:TBA_P_pseudoE} we find an implicit solution for the dressed energy
\begin{equation}
E(\ell_s^2,R) = \frac{1}{\bar{R}}\left(R  + \frac{\ell_s^2}{2}E(\ell_s^2,R)\right)E(0,\bar{R}) + \frac{\ell_s^2}{2\bar{R}}P(R)P(\bar{R})
\label{implicitE_pnonzero}
\end{equation}
which is the solution  of the hydrodynamical dressing equation obtained in \cite{Cavaglia:2016oda}.

\section{Leading order multiparticle amplitudes with an arbitrary number of legs}
\label{PS}
In this appendix, following \cite{Dubovsky:2016cog},  we show how to obtain the generating functional for leading inelastic amplitudes  \eqref{3Dundlag} (with an arbitrary number of legs) from the Polchinski-Strominger formalism. 

 Following \cite{Hellerman:2014cba}, the main ingredients of the PS formalism are, as in the standard Polyakov construction, the worldsheet metric $h_{\alpha\beta}$ and the embedding coordinates $X^{\mu}$. For simplicity we restrict to the $D=3$ case, although
 not much changes at any other value of $D$. One can introduce the composite Liouville field
\begin{equation}
\phi = -\frac{1}{2}\log\left(h^{\alpha\beta}\partial_\alpha X^\mu\partial_\beta X_\mu\right)
\end{equation}
which under Weyl transformations $h_{\alpha\beta} \to e^{2\omega}h_{\alpha\beta}$ transforms as
\begin{equation}
\phi \to \phi +\omega.
\end{equation}
Using this composite operator one introduces the PS term in the action which allows to fix the central charge to the critical value $c=26$ without introducing new degrees of freedom. As a result, the theory enjoys Weyl gauge invariance even in the non-critical case.
With the help of $\phi$ one can construct the  Weyl covariant derivative $\hat{\nabla}$, which acts on covectors $A_\beta$ as

\begin{equation}
\hat{\nabla}_\alpha A_\beta \equiv \nabla_\alpha A_\beta + \nabla_\beta A_\alpha - h_{\alpha\beta}\partial^\gamma A_\gamma\;.
\end{equation} 
A general  action is constructed by writing all possible terms invariant under diffeomorphisms and Weyl transformations using these ingredients. 

 Let us focus on the leading non-universal interaction, which is of the form
\begin{equation}
\label{Sint}
S_{int} = \frac{\gamma}{4}\int d^2\sigma \frac{\left(h^{\alpha\alpha'}h^{\beta\beta'}\hat{\nabla}_\alpha\partial_\beta X^{\mu}\hat{\nabla}_{\alpha'}\partial_{\beta'} X_{\mu}\right)^2}{\left(\partial_\gamma X^{\nu}\partial^\gamma X_\nu\right)^3}\;.
\end{equation}
To obtain  tree level amplitudes in the undressed theory one first makes use of the gauge symmetries (diffeomorphisms and Weyl) to fix the Polyakov metric to be flat.  The gauge fixed action has to be supplemented with  the  Virasoro constraints, which at leading order in $\gamma$ take the free form
\begin{equation}
\partial_\pm X^\mu\partial_\pm X_\mu = 0
\end{equation}
Expanding around the flat worldsheet background
\begin{align}
X^\pm \equiv \frac{X^{0}\pm X^1}{2}& = \frac{\sigma^\pm}{\ell_s} + Y^\pm\\
X^2 &= X
\label{flatPS}
\end{align}
one gets
\begin{equation}
\partial_\pm Y^\mp = \frac{\ell_s}{2}\left(\partial_\pm X\right)^2.
\label{solveVir}
\end{equation}
To obtain the generating functional for leading order undressed amplitudes one simply plugs in this solution in the action (\ref{Sint}). This gives (on-shell)
\be
\label{gen}
 S_{int}={ 32 \gamma} \int d^2\sigma {(\partial_+^2 X)^2(\partial^2_- X)^2\over( 4-\ell_s^2 \partial_+X\partial_-X)^6}\;.
\ee
Expression (\ref{gen}) is the interacting part  of the undressed action (the dressing comes about by switching from the worldsheet coordinates $\sigma^\pm$ to the physical ones $X^\pm$ \cite{Dubovsky:2016cog,Dubovsky:2017cnj}). Unlike in the full dressed theory, 
at the leading order in $\gamma$ each amplitude following from (\ref{gen}) is given by a single tree level vertex, so that (\ref{gen}) can be also considered as a generating functional for leading order undressed amplitudes.
We see that the PS formalism provides a very convenient way to account for  dressing of any higher-dimensional operator by the Nambu--Goto vertices. Expanding (\ref{gen}) up to sextic order in fields one reproduces (\ref{3Dundlag}).

\bibliographystyle{utphys}
\bibliography{bibliography3partftube}

\end{document}